# INFLUENCE OF SOLAR ACTIVITY ON STATE OF WHEAT MARKET IN MEDIEVAL ENGLAND

Lev A. Pustilnik (1), Gregory Yom Din (2)

(1) Israel Cosmic Ray Center, Tel Aviv University &Israel Space Agency, Israel;
(2) Golan Research Institute, Kazrin, Israel

ABSTRACT.   The database of Prof. Rogers (1887), which includes wheat prices in England in the Middle Ages, was used to search for a possible influence of solar activity on the wheat market. We present a conceptual model of possible modes for sensitivity of wheat prices to weather conditions, caused by solar cycle variations, and compare expected price fluctuations with price variations recorded in medieval England.

We compared statistical properties of the intervals between wheat price bursts during years 1249-1703 with statistical properties of the intervals between minimums of solar cycles during years 1700-2000.  We show that statistical properties of these two samples are similar, both for characteristics of the distributions and for histograms of the distributions. We analyze a direct link between wheat prices and solar activity in the 17th Century, for which wheat prices and solar activity data (derived from 10Be isotope) are available. We show that for all 10 time moments of the solar activity minimums the observed prices were higher than prices for the correspondent time moments of maximal solar activity (100% sign correlation, on a significance level < 0.2%). We consider these results as a direct evidence of the causal connection between wheat prices bursts and solar activity.

## 1. Introduction: Connection between Solar Activity and Terrestrial Environment

A search of the influence of solar activity on the Earth reveals several periods characterized by radical reversals in scientific opinions. There was an early period, from the Middle Ages ("Astrological" period) and from the 19[th] to the beginning of the 20[th] Centuries ("Romantic" period), when numerous enthusiasts accepted the influence of solar activity on the terrestrial environment as an evident fact that did not need any critical analysis. After this, a "Pessimistic period" started from quantitative measurements of the real variability of solar energy output detected on the Earth's orbit. These measurements showed that the main parameter of solar influence, emission input into the atmosphere, is extremely stable, as reflected in the name given to the phenomenon – "the solar constant".  According to the latest measurements, its variability, caused by sunspot cycle, is limited by an amplitude $A < 0.1\%$ (Frochlich and Lean, 1997). This limit leads to correspondent variations in the Earth's temperature ($\Delta T < 0.1°$) too small to allow the observation of essential manifestations in the terrestrial environment.

However, in recent times another parameter of solar activity was discovered, with a very strong variation in time, on a time-scale of variability from ten to hundreds of years. This parameter is the magnetic activity caused by the emergence of magnetic fields from the convective zone to the photosphere and the external solar atmosphere. Over the past 30 years, it has been understood that this magnetic drive is a dominant factor in coronal gas heating and in the formation of solar winds.

We will not discuss here the latest advances in dynamo-models of the solar magnetic field, but will summarize only several points, relevant to our present considerations.

The generation of the solar magnetic field has a cyclical character. According to measurement of the effects, dependent on the absolute value of the magnetic field (sunspots, photosphere bright regions (flames), coronal X- and radio emissions), this cycle is not a regular process, but changes over periods varying from 8 to 17 years (Fig.1a), with an average solar cycle period of about 11 years. The value of the best-known parameter of the solar cycle, the so-called "solar spot number" or "Wolf number", varies during the cycle from 0 to 100-200.

A natural consequence of the instability of the solar cycle period is a phase deviation. The phase of the minimum point, for example, shows long-term oscillation of up to half a period, up and down (Fig .1b).

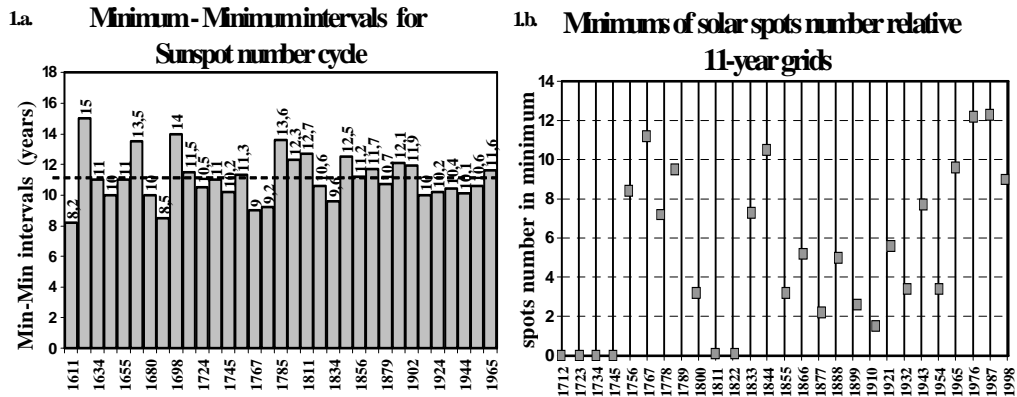

*Figure 1. (a)* Illustration of the instability of the solar cycle period: variation of intervals between minimums of the solar spot cycles over 1800-2000 years (X-axis is year of minimum; Y-axis is a time interval up to the next minimum). As it can be seen, cycle duration varies from 8 – 15 years. The dashed line corresponds to the average 11-year period. (*b*) Illustration of a phase deviation: moments of solar spot minimums in the 11-year grid-net demonstrate a permanent slow deviation of phase of solar spot minimums relative to the phase of the 11-year cycle.

We would like to point here the fact, essential for the next part of this paper, that the absence of the dark spots on the solar surface in the minimal phases does not mean a full termination of solar activity. As it was shown by researchers of solar spot formation in the photosphere, solar spots become visible only when the magnetic field in the spots exceeds a critical value (about 2000 Gs), necessary to prevent convection motion and cooling of plasma at the photosphere level. For magnetic fields below the critical level, no dark spots will be formed, and optical observation in white light will demonstrate a "zero" level of activity, despite magnetic activity that is, in fact, manifested directly in both coronal and solar wind activity.

In spite of a very low energy output from coronal activity and solar wind, their influence on the terrestrial environment is very strong. For our purposes, the most essential aspect of this influence is the modulation of the galactic cosmic ray (CR) flux, penetrating from the Galaxy into the solar heliosphere. This penetration is a result of two opposite processes: the ejection of galactic CR by the magnetic field of the solar wind, and diffusion of CR in the heliosphere cavity. The dynamic equilibrium between these processes determines the level of the cosmic ray flux entering into the Earth's atmosphere. The strong dependence of solar wind on solar activity leads to corresponding variations in the cosmic ray flux into the Earth atmosphere caused by the solar cycle.

On the other hand, as understood recently, cosmic rays are an influential factor on terrestrial cloud formation, since cosmic ray particles are an essential source of atmospheric ionization and radicals formation. Atmospheric ions and radicals, created by these high-energy particles, form numerous nuclei in the atmosphere, necessary for the process of the condensation of water vapor that leads to the formation of clouds. As recently shown by Svensmark and Friis-Christinsen (1997), there is a strong connection between the cosmic ray level and cloudiness of the Earth's atmosphere (Fig 2a.). This discovery restored a physical base for a possible understanding of the influence of "space weather" on terrestrial weather, and stimulated numerous attempts to correlate these phenomena. It was shown that sensitivity of the Earth atmosphere to cosmic ray variation depends strongly on the coordinate on the Earth surface and atmospheric level height.

In addition to the well-established 11-year cycle of solar activity, a long-term variation of the amplitude of solar cycle has been established (Friis-Christensen and Lassen, 1991). In these variations, both in the maximal number of solar spots and in the duration of the cycles, the latter parameter shows a correlation with mean Earth temperatures. The so-called Minimums of Wolf, Sperer, Maunder and Dalton belong to the most drastic of these long-term variations, when manifestations of solar cycle activity were minimal over many decades. The best known of these "break-offs" is the Maunder Minimum (1645-1715), which took place when the telescope came into wider use, after the first observations by Galileo in 1614. The absence of solar spots during the Maunder minimum was attested by numerous enthusiasts of astronomy: observational time covered more than 90% of the Maunder Minimum time. As a result, reliable data for analysis was obtained. A comparison of temperatures in that period in northern Europe, the Atlantic and Greenland has led researchers to the conclusion that temperatures during the Maunder Minimum in Europe were significantly lower than before and after this period (Fagan, 2000). This led to a suggestion of a possible causal connection between decreased solar activity and weather conditions during that period. Different methods have been used to establish this connection, or other such effects on the Earth's environment.

1. A search for any direct correlation between the number of sunspots (or solar cycle duration) and observed parameters in the Earth's environment (temperature, atmospheric pressure, ozone column density, radiation input). A possible time lag between solar activity and its influence created problems for this approach, especially when this lag was variable. As a result, it was shown that the observed effects depended on: a) geographical location – signs of correlation and amplitude change from place to place on the Earth; b) epoch of observation - signs of correlation can change from one long-term time period to another.

2. A search of 11-year harmonics in frequency spectra of Earth effects (temperature, pressure, ionization level). Unfortunately, solar activity is not a good object for this method since: a) solar activity periods are not stable, but vary at wide intervals (Fig.1a.); b) The time lag between solar activity and terrestrial weather can vary.

As the result of these inconsistencies in the solar cycle phase, investigators of connections between solar and terrestrial weather were forced to divide the time interval of research in a number of short sub-intervals and to estimate connectivity for each sub-interval separately. Sometimes this connection demonstrates changes in the signs of the correlation from one sub-interval to another. Since data on connectivity between Sun and Earth weather obtained by this method are not reliable enough, the existence of this connection is not widely recognized.

Hence, it is necessary to search for other possible forms of response observed on the Earth, much more sensitive to the solar cycle or to its selected phases. In the present study, we consider wheat prices in the medieval market in England, with its limited sources of supply and relative isolation from continental markets, as such a possible response.

## 2. Earlier efforts to find the influence of solar activity on wheat prices in England

The first suggestion of a connection between wheat price and sunspots was made by William Hershel (1801). Over 40 years (1779 – 1818), Hershel regularly observed sunspots and their variations in number, form and size. Unfortunately, most of his observations took place in a period characterized by the lowest solar activity since the Maunder Minimum (later called the Dalton Minimum), when solar activity behaved very unusually: spots in minimums disappeared totally, and the max-max interval increased up to 17 years (1788-1805). These irregular variations of sunspot numbers prevented Hershel from discovering the 11-year period in solar activity. However, he paid attention to an evident correlation between the observed number of spots and the state of the wheat market, based on a series of wheat prices published by Adam Smith in his classical work "Wealth of Nations" (1776). As Herschel showed in his report to the Royal Society (1801), five prolonged periods of few sunspots correlated with costly wheat. In 1801, Hershel published the results of his observations of sunspots, remarking on the possible connection between sunspots and wheat prices, in the Philosophical Transactions of the Royal Society. Unfortunately, his conclusions met with such strong and widespread rejection, even ridicule[1], that in spite of his great authority and his status as Royal Astronomer, he had to cancel his next public presentations of his results. However, it is known from his biography, that he continued sunspot observations and analysis up to last days of his life. Hershel suggested that a possible reason for the influence of sunspots on wheat prices was changes in the Earth climate produced by modulations in solar radiation, caused by variation of the emitted surface[2].

The next scientist in this field was the well-known English economist and logician William Stanley Jevons, one of the creators of Neoclassical Economic Theory. In his study (1875), he focused on the first part of the data published by Professor Rogers in the first volume of his work. Wheat prices over 140 years, from 1259 to 1400, were presented in this volume. Jevons discovered that the time intervals between high prices were close to 10-11 years.

The coincidence of these intervals with the period of the recently discovered 11-year cycle of solar activity led him to suggest a solar cycle as a "synchronization" factor for fluctuations in wheat prices[3] (Jevons, 1878). As a next

---

[1] *Lord Brougham ridiculed them as a "grand absurdity."*
[2] *The first author of this idea was the famous Jonathan Swift, who described in "Gulliver's Travels" (1726) Laputian scientists worried that "the Face of the Sun will by degrees be encrusted with its own Effluvia, and give no more light to the World".*
[3] *In 1875 and 1878, Jevons read two papers before the British Association which expounded his famous "sunspot theory" of the business cycle. Digging through mountains of statistics of economic and meteorological data, Jevons argued that there was a connection between the timing of commercial crises and the solar cycle. The basic chain of events was that variations in sunspots affect the power of the sun's rays, influencing the bountifulness of harvests and thus the price of corn which, in turn, affected business confidence and gave rise to commercial crises. Jevons changed his story several times (e.g. he replaced his European harvest-price-crisis logic with an Indian harvest-imports-crisis channel). In spite of the weakness of his explanation, Jevons believed that the periodicity of the solar cycle and commercial crises is approximately 10.5 year by his calculations. It was too coincidental to be dismissed. This verged on the eccentric, and furthermore, the statistics did not bear him out. Nonetheless, it remains a significant piece of work, as this was perhaps the first time that the phenomenon of a business cycle was identified. Economists had long been aware that business activity had its ups and downs, but not that they necessarily followed any regular pattern. They generally*

step, he extrapolated his theory to stock markets of the 19th Century in England and was impressed by a close coincidence of five stock exchange panics with five minimums in solar spot numbers that preceded these panics. He suggested that both solar and economic activities are subjected to a harmonic process with the same constant period equal to 10.86 year. However, the subsequent discovery of the non-harmonic behavior of solar cycles, with periods varying from 8 to 15 years, and the later observation of lack of coincidence between panics predicted by Jevons and actual ones, destroyed his arguments. A notable statement by of one of his critics, Prof. Proctor (1880), notes that, under conditions of variable periods of solar activity, it would be more effective not to use the ephemeral "harmonic period", but to confront a chosen phase of solar activity with the state of the market, and to search for any correlation between them.

3. **Conceptual model of possible influence of solar activity on price dynamics**

The physical basis of the influence of solar activity on the terrestrial climate is quite evident. High-energy cosmic rays penetrate the atmosphere to create large numbers of ions and radicals – centers for the condensation of water vapor and the formation of clouds. Increases in the number of sunspots and magnetic solar activity lead to an increase in solar wind and to a stronger ejection of galactic cosmic rays from the solar system. Under these conditions, with maximum solar activity and strong solar wind, cosmic ray radiation on the Earth's orbit decreases, depressing ion and radical formation in the atmosphere. The effect on the climate is decreased cloudiness. Clear skies, increased temperatures and a decrease in precipitation result in hot weather and drought. On the other hand, a decrease in the number of sunspots and magnetic solar activity leads to a decrease in the solar wind and to an increase in cosmic ray radiation on the Earth's orbit. The density of ions and radicals in the Earth's atmosphere increases, resulting in higher formation of clouds and to higher rain/snow precipitation. Clearly, a long-term breakdown of solar activity over tens of years (e.g., the Minimums of Maunder, Sperer, etc.) will lead to the effects of a minor Ice Age.

A possible influence of space weather on the behavior of wheat prices can be multiplied by nonlinear links in the causal chain: "solar activity" – "climate" – "wheat production" – "market prices". This nonlinearity can lead to major variations in the output parameter (price), as a result of relatively small changes in the input parameters. Some of these possible nonlinear links are described below.

Recent data from the satellite ISSP-D2 (Rossow and Shiffer, 1991; Fastrup *et al.*, 2000) demonstrate a direct link between cloudiness and cosmic ray flux, modulated by solar activity (Fig.2a). We note that these data gave an estimate of a global effect, averaged over a large territory, in terrestrial terms. At the same time, the detailed data show a much more complex picture: the correlation of cosmic ray flux on the Earth's surface with weather characteristics is not uniformly distributed (Fastrup *et al.*, 2000), but includes numerous spots of high positive and high negative correlations, both as regards cloud cover and cloud temperature (Fig.2b). For example, in England over the last 15 years, there is a specific zone with a high positive correlation in cloud cover and a high negative correlation with cloud temperature.

---

*believed that "crises" arrived haphazardly, punctuating the smooth advance of the economy at irregular intervals. Jevons was perhaps the first economist to argue that the phases of business activity had a regular, measurable and predictable periodicity (http://cepa.newschool.edu/het/profiles/jevons.htm).*

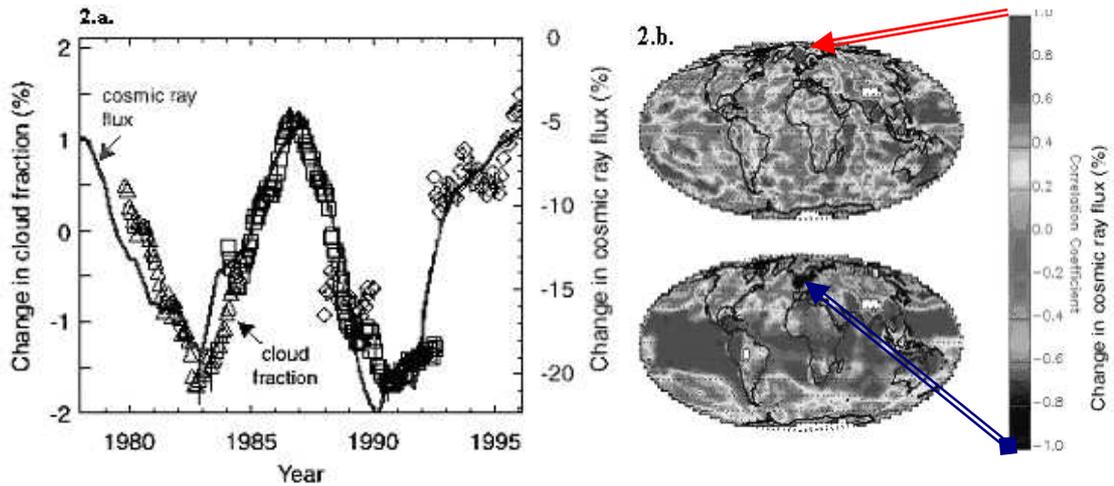

Figure 2. (a) Cloudiness, averaged over the Northern Hemisphere, and simultaneous CR flux variation during 1978-1996. (b) Distribution of the correlation coefficient (r=1 – "white", r=-1 "black") over Earth. The upper map is for cloud fracture (surface), the bottom map is for cloud IR temperature. The England region (arrows) is a place, where CR have high positive correlation with cloud fracture and high negative correlation with cloud temperature – an increase of cosmic ray leads to an increase of cloudiness and a decrease in temperature (Fastrup et al., 2000).

Variations in global cloudiness caused by solar cycles are not high – only a few percent. Hence, in a linear system, correspondingly small responses in temperature, wheat production and price could be expected. However, two factors can lead to high sensitivity of the output parameter (market price) to initial variations in solar activity.

First, in comparison with effects averaged over the Northern Hemisphere, effects observed in small regions with a high "cosmic ray"-"weather" correlation must be much stronger and appear as local weather anomalies correlated with maximum or minimum of solar activity, such as long periods of heavy rain, extreme frosts and severe winter, on the one hand, and torrid summer with severe drought, on the other.

Second, the link between wheat production and price in a market with a limited supply (as it was in medieval England) is nonlinear, and has critical marginal zones, where relatively small variations in wheat production can lead to sharp changes in prices, similar to phase transition. In detail, the nonlinear sensitivity of the output parameter to solar activity and variations in cloudiness can be characterized by:

a. High variations in grain production, as a result of only a few abnormally cold days during germination, or an additional week of rainfall or, on the contrary, of drought, at critical periods (harvest, foliation). At the same time, even small changes in mean annual temperature can cause major changes in the geographical location and area of crops cultivated[4].

b. The nonlinear sensitivity of prices (in a wheat market with limited supply) to any shortage or excess of wheat supply. These effects can lead either to a catastrophic rise in prices, in the case of shortage, or to a sharp drop, in the case of overproduction.

These evident and expected non-linearity opens new possibilities for a search

---

[4] *From Carter et al. (1992): "… a mean annual temperature increase of only 1ºC would open up large areas in southern England, the Low Countries, eastern Denmark, northern Germany, and northern Poland to potential grain maize cultivation. An increase of 4ºC would move the limit into central Fennoscandia and northern Russia."*

of possible connections between wheat prices and solar activity.

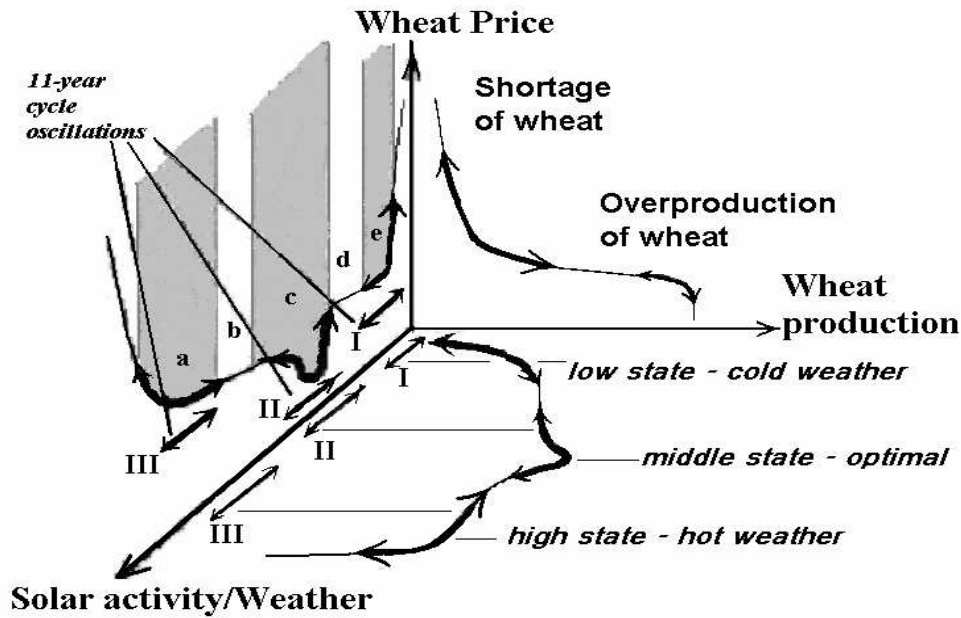

*Figure 3. Conceptual model for wheat market in medieval England - a three-dimensional diagram of the possible influence of solar cycle/weather conditions on wheat prices. The state of the system is defined by its position in the 3-D space ("Solar activity/Weather"-"Wheat Production"-"Wheat Price"). The amplitude and the sign of the influence depend on the state of the system in a weather-wheat production plane (low, middle, high) and on a range of weather fluctuations caused by solar cycle variability (size of arrows)*

We illustrate this concept in Fig. 3, which describes the system state as a point in a three-dimensional space: "solar activity/weather"-"wheat production"-"wheat price". On the plane of "solar activity/weather"-"wheat production", the position on the "weather" axis is the result of a superposition of two processes: slow weather change caused by the terrestrial climate evolution and fast (11 +/- 3 years) oscillations in cloudiness, temperature and rain, caused by solar cycle variation of cosmic ray input (shown by arrows). The reaction of wheat production to the fast variations depends on the position of center of the oscillation interval relative to the threshold of nonlinear sensitivity (**a, b, c, d, e** – respectively). If, in the result of slow climate evolution, the center of an interval of weather oscillation initiated by the solar cycle shifts to a boundary of the zone of high-risk agriculture ("**e**" - "low state – cold weather" state or "**a**" - "high state – hot weather ") it can lead to periodic shortages of wheat, caused by weather anomalies, such as droughts for the "hot weather" state, or extended rainfalls and severe frosts for the "cold weather" state. Clearly,, the decrease in wheat production in "lean years" will lead to an extreme rise in prices. Another situation occurs when the center of an interval of weather oscillation caused by a solar cycle shifts to the zone **"c"** of optimal conditions for wheat production. This will lead periodically to "fat years", with an overproduction of wheat and a consequent collapse in prices.

As presented on the plane "Production"-"Price", a shortage in wheat in "cold" or "hot" zones of the axis "Solar activity/Weather", or overproduction in an "optimal" zone will lead to extreme rises in price in minimal solar activity years ("cold" zone) or in maximal activity years ("hot" zone), and to low prices for an "optimal" zone. On the other hand, it is evident that for zones **"b"** and **"d",** where the nonlinear effect of solar activity on wheat production is minimal, variations in wheat production and prices, correlated with solar activity, are not expected.

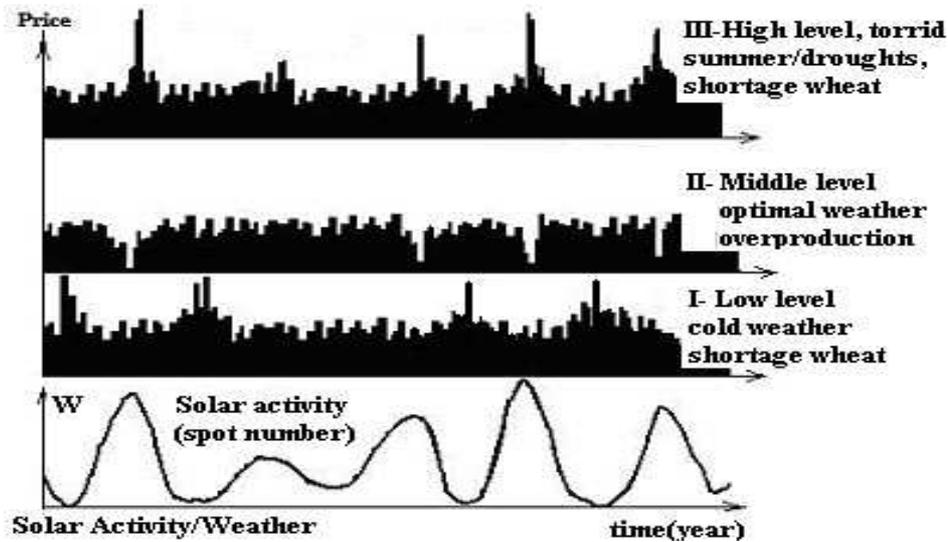

*Figure 4. Expected types of price variation caused by solar activity cycles. I – expected price bursts for the "Low state" sensitive to cold weather/cloudiness leading to shortage in wheat supply, II – prices for the "Middle state" with regular generation of optimal conditions for wheat production, leading to price fall down caused by "overproduction", III – prices for the "High state" sensitive to hot weather – a decrease in cloudiness leading to a shortage in the wheat supply.*

Three scenarios of possible reactions of the wheat market to variations in the solar activity cycle for three states on the "Solar activity/Weather" axis ("cold", "optimal" and "hot" in terms of Fig. 3) are presented in Fig. 4. As shown, price bursts of different kinds are possible for these states: 1) bursts in the maximum solar activity (case "III") for a "hot" state caused by heat and drought; 2) bursts in the minimum solar activity (case "I") for a "cold" state, caused by anomalous severe winter or rains; 3) for case "II" with an "optimal" state for wheat production we can expect troughs in prices, when weather variations led to warm weather by enfeebling normal "severe winter" extremes for cold places (in the maximum of solar activity) or by weakening normal torrid summer/drought extremes for hot places (in the minimum of solar activity). Clearly, extreme bursts in prices can take place especially in markets with a low level of globalization, when a shortage in local production of wheat cannot be compensated by supply from other regions.

    In the case of medieval England, with its isolated wheat market and severe limitations on an external wheat supply, we suggest, that a price burst would be the reaction to the solar activity cycle. Wheat production in the medieval England was under conditions of "high-risk" agriculture of the "cold' type (negative influence of severe winters and long-time rains). Hence, we must expect positive burst reactions in the years of minimal solar activity, when minimal solar wind and maximum cosmic ray input into the Earth's atmosphere caused additional cloud production and a temperature decrease (see the positive correlation of the cosmic ray variations with cloudiness, and negative correlation with cloud temperature in the England region, Fig. 2). On the other hand, we must take into account that long-term global climate variations could lead, at certain times, to a "favorable" weather state in the England region, with low sensitivity to the influence of the cloudiness variation caused by solar activity cycle. The expected price bursts for "boundary" weather states

requires us to use relevant methods in the search for correlation, methods developed to investigate burst-like events. Based on this approach, we will test two properties of the price bursts:

a. The distribution of the time intervals between high bursts, and its coincidence (or not) with the distribution and phase curves for extremes of solar cycle phases (minimal or maximal number of sunspots).

b. Direct correspondence between the price level and a solar activity phase for the selected time interval, when both price data and solar activity data are available.

### 4. Analysis of data

Three samples of data were used for our analysis:
- Agricultural prices **P(t)** in England for 1259-1702 from Prof. Rogers (1887).
- Moments of maximums/minimums of sunspots [3] $T_{Max}(t)$, $T_{Min}(t)$ for years 1610 - 2000 from NOAO Satellite and Information Center at ftp://ftp.ngdc.noaa.gov/STP/SOLAR_DATA/SUNSPOT_NUMBERS/maxmin.new
- Data on solar activity maximums from isotope $^{10}Be$ in Greenland for the years 1600-1700 from Beer *et al*. (1998) and Usoskin *et al*. (2001).

#### 4.1. SAMPLE FOR ANALYSIS

As set out earlier in this work, price bursts can serve as an effective manifestation of connection between solar activity, cosmic rays and terrestrial weather. Hence, we began our analysis by creating a sample of price bursts from data of yearly wheat prices (Rogers[4], 1887). The graph of the time-series of wheat prices (Fig. 5) contains two specific features:

1. A transition from "low price" level to "high price" level during 1530–1610, possibly due to access to sources of cheap silver in the recently discovered New World. We use a logistic curve for fitting our data, aiming at the long-term extrapolation of this transition.

2. The existence of two types in the price variations: noise-like variations and several high amplitude bursts.

---

[3] Solar spots were observed regularly from their discovery by Galileo in 1612, but as a result of the Maunder Minimum (1645-1735), when spots almost disappeared, sunspots were detected rarely and irregularly. Only after 1700, when spots came back to the face of the Sun and regular solar spot observations by several observers were restored, can we consider the observed data as reliable. Unfortunately, the Maunder Minimum covers a major part of the period (1612-1702) when data on both prices and sunspots are available. This forces us to use indirect parameters (isotopes $^{10}Be$, $^{14}C$, sensitive to cosmic ray flux) to compare prices and solar activity during this period.

[4] The wheat market in Medieval England is a good source of data for examining manifestations of solar activity through wheat prices. Agriculture in England of this period was very sensitive to weather conditions (with catastrophic drops in grain production as the result of abnormally severe winters, or insufficient solar radiance during the growth period). Our source of data on prices is the 50-year research of Prof. Rogers, who collected data on agricultural prices for the 450-year period (1259–1702). Roger's data were drawn principally from the account books of numerous English monasteries, and in part from those of landowners. Clearly, these data contain minimal subjective distortions and preferences. English monks considered their work as a mission, and were therefore very responsible observers of nature (like their Scandinavian colleagues that collected data of auroras). The wheat market in Medieval England is an example of a market with a limited supply - England being an island.

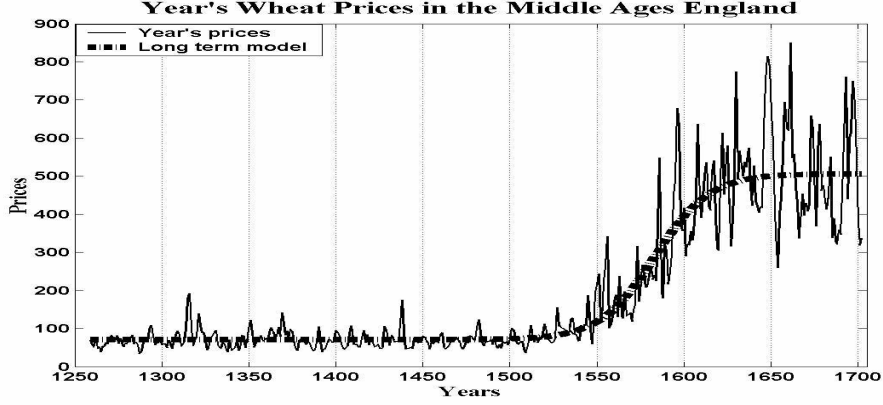

*Figure 5.* Wheat prices in England 1259-1702 with price transition 1530-1610.

We compared the distribution of intervals of price bursts with the distribution of the intervals between extremes (minimum phases) of solar cycles.

Our first step was to estimate parameters of the logistic curve as a long-term model $\hat{P}(t_i)$ for the transition of the prices $P(t_i)$ from "low" to "high". Deviations in annual prices, from year to year, were calculated as $\Delta P(t_i) = P(t_{i+1}) - P(t_i)$, normalized in respect to the value $\hat{P}(t_i)$ of the long-term model: $\delta P(t_i) = \Delta P(t_i)/\hat{P}(t_i)$, reflecting price bursts. We used Absolute values of the Normalized Deviations of Prices: $ANDP(t_i) = Abs(\delta P(t_i))$, which are sensitive not only to sharp increases, but also to sharp falls.

The second step of our analysis was to choose a discrimination level, filtering out low amplitude bursts, generated by the noise component of the variation. This enabled us to select only anomalous price bursts, potentially connected with the space weather and solar cycles (evidently, in the result of the filtering we can lose a part of the real low amplitude bursts). For our data, a discrimination level $ANDP(t) \approx 50\%$ was chosen (Fig. 6).

The third step was to divide the interval of 443 years (1259-1702) by the 11-year grid and to identify points of maximum normalized price deviation *ANDP* in the inter-grid intervals as points of price bursts (marked by triangles in Fig. 6). Two segments of the sample, 1313-1324 and 1544-1556, marked by white rectangles, were not included in the analysis, because during these periods very strong fluctuations of price from year to year took place (possibly caused by political or military disturbances). Bursts and their times are presented in Table 1.

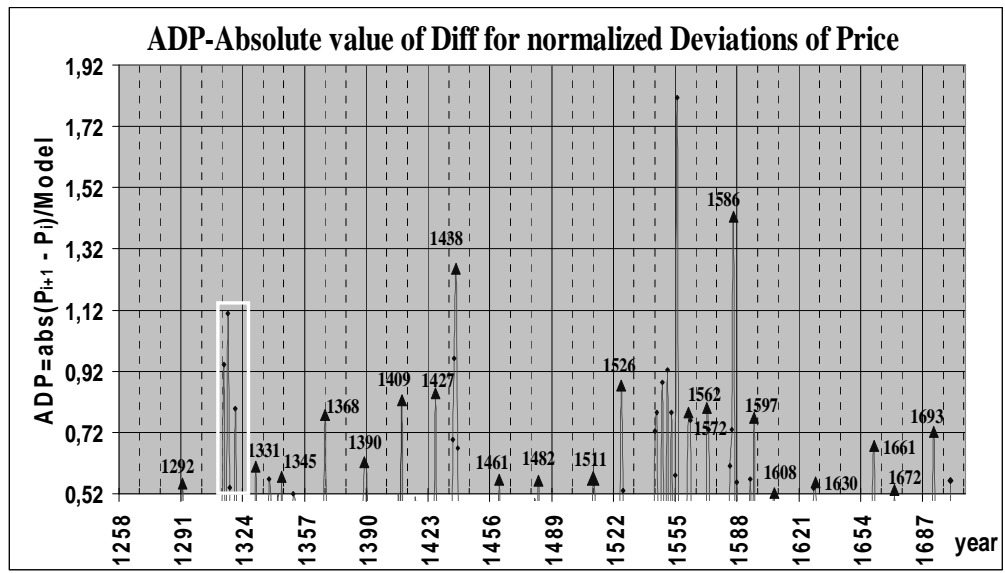

*Figure 6.* Absolute values of normalized relative deviations of annual prices (ANDP). Vertical grids on X-axis mark 11-year intervals. The triangles mark price bursts maximal for the given 11-year interval; labels near marks are years of bursts.

4.2. INTERVAL ANALYSIS

Times of price bursts are used to generate a sample of intervals between them, which are clearly divided into two groups: the first with intervals lasting about 10-11 years, and the second intervals that are multiples of 10-11 years. We consider this to be the result of losing some bursts, due to their low amplitude. Hence, for this second group we restored possible original values by dividing of burst- burst time intervals by the corresponding number of the intervals between bursts (2, 3, …) in the 11-year grid (Fig. 6.).

The main question for our test is whether the interval distribution for price peaks is statistically similar to the distribution of intervals of extremes of solar activity (for example, minimum-minimum intervals in sunspot variations). In Fig. 7, we present two histograms: a) for restored burst-burst intervals of prices over 450 years (1249-1702), based on Table 1, the last two rows; b) for intervals between minimums in solar activity cycles during the last 300 years (1700-2000), based on the NOAO sample of solar maximums/minimums $T_{Max}(t)$, $T_{Min}(t)$.

Conclusions from the interval analysis:
1. For the distribution of sunspot minimum-minimum intervals the estimated parameters are: median – 10.7 years; mean – 11.02 years; standard deviation – 1.53 years.
2. For the distribution of price burst intervals the estimated parameters are: median – 11.0 years; mean – 11.14 years; and standard deviation – 1.44 years.
3. The null hypothesis that the frequency distributions are the same for both of the samples (intervals between price bursts and intervals between minimums of sunspots) can not be rejected with $\chi^2$-test (significance level >95%).

*Table 1.*
*Time of the price bursts, number of cycles, average burst-burst intervals*

| Year of bursts | Number of cycles | Interval |
|---|---|---|
| 1292 | 4 | 9.75 |
| 1331 | 1 | 14 |
| 1345 | 2 | 11.5 |
| 1368 | 2 | 11 |
| 1390 | 2 | 9.5 |
| 1409 | 2 | 9 |
| 1427 | 1 | 11 |
| 1438 | 2 | 11.5 |
| 1461 | 2 | 10.5 |
| 1482 | 2 | 14.5 |
| 1511 | 1 | 15 |
| 1526 | 3 | 12 |
| 1562 | 1 | 10 |
| 1572 | 1 | 14 |
| 1586 | 1 | 11 |
| 1597 | 1 | 11 |
| 1608 | 2 | 11 |
| 1630 | 3 | 10.3 |
| 1661 | 1 | 11 |
| 1672 | 2 | 10.5 |
| 1693 | | |

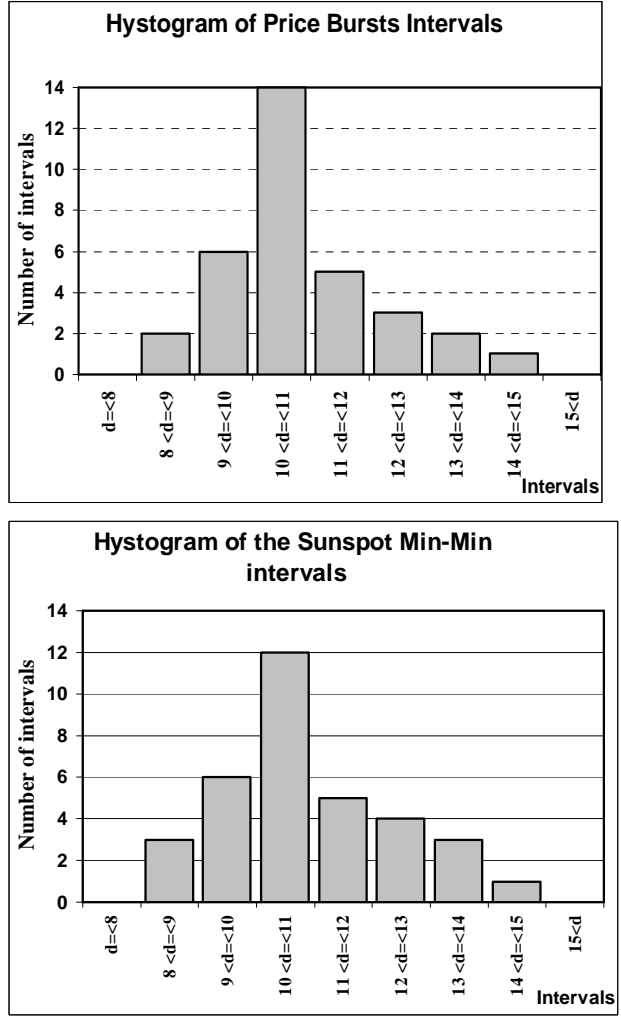

**Figure 7.** Histograms of the interval distribution for price bursts and for minimum-minimum intervals of sunspots

### 4.3. PHASE ANALYSIS

The next property to be tested is the phase distribution of the price burst moments. This analysis is based on the suggestion that some harmonic process takes place in price bursts and solar activity. It is understood that the instability of the solar cycle period and its slow phase deviations, demonstrated earlier in our work, makes this test less efficient. In reality, if during the observational time a given period of the cycle dominates, the phase of the chosen state (of price burst or of number of sunspots) must remain the same. The drift of the period or its slow oscillation will scatter a phase value relative to the dominant one, but some significant excess will be detectable. Hence, the phase distribution of the chosen state can show a reliable maximum, if the phase drift or a variation of cycle periods does not suppress it. A phase of the moment of time $t_i$ is determined as:

$$\Phi_i = (t_i - t_0)/T_0 - \text{int}\left[(t_i - t_0)/T_0\right],$$

where **int** is an integer part of the value, **T₀** is a period of the cycle and **t₀** is a zero point of the cycle (for our aims this parameter is not essential and we will use **t₀** = 0). In this test, we suggest that our sample of price bursts is a manifestation of some cyclical process with an unstable period, such as the solar cycle. If this hypothesis is correct, the same phases will be obtained for most of the bursts; if it is false, the phase distribution of the bursts will be close to uniform.

Fig. 8 shows the phase distributions of price bursts for a period of 11.14 years (we use the estimated mean value for price burst-burst intervals for the period of 1249-1703) and for moments of minimal solar activity for a cycle period of 11.02 years (mean for minimum-minimum solar spot intervals for the period of 1610-1986).

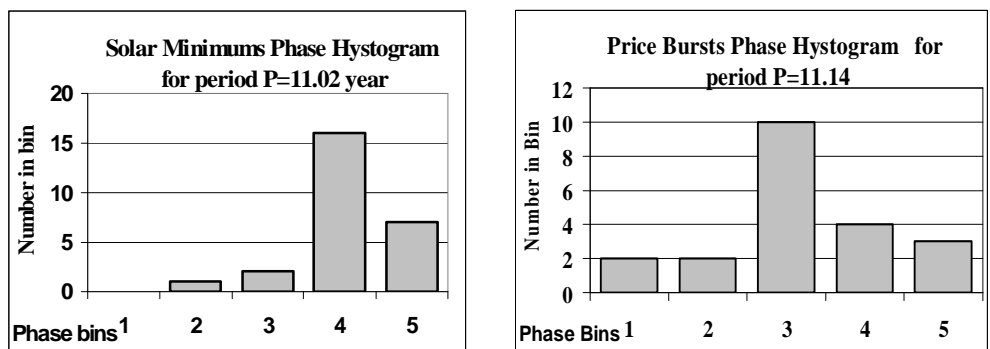

**Figure 8. a.** Phase distribution for price bursts (for period 11.14 years); **b.** Phase distribution for solar activity minimums (for period 11.02 years).

The hypothesis that the distribution of the phases is uniform can be rejected on a significance level >99% for price bursts and >99.9% for minimums of solar activity. This result indicates the existence of the mean cycle period for price bursts and for solar spot minimums, with conservation of the accumulated phase in spite of essential phase variations which occur from time to time

The interval and phase analysis for price bursts shows that high amplitude price bursts reflect the manifestation of some cyclical process with a period of about 10-11 years, close to the solar cycle period. Statistical characteristics are not distinguishable in either processes (price bursts and solar activity), on high level of confidence. We note here that this result does not prove any causal connections between solar cycle and price bursts, because this may be a random coincidence (similar, for instance to the closeness between the values of the modern solar cycle, 11.02 years, and the orbital period of Jupiter, 11.86 years). Hence, to prove a causal connection between the solar cycle and the wheat price level, we must find a systematic correlation between any extreme phase of solar activity (maximum or minimum) and the correspondent price level, for any chosen time interval.

### 5. Correspondence between price level and solar cycle phase

The main problem for a comparison between price levels and solar activity is the absence of a time interval common to reliable sunspot observation data (1700-2001 years) and wheat price data (1259-1702 years). Nevertheless, the period 1614-1702, when astronomers already started to observe solar spots, can be used for this comparison. Unfortunately, most of this period is covered by the Maunder Minimum, when solar activity was so strongly depressed that spot manifestations were absent or very weak, irregular and disturbed. However, the discovery of the strong anti-correlation between concentration of $^{10}$Be isotopes in Greenland ice, generated by cosmic rays and solar spot activity (Beer *et al.*, 1998), opens new possibilities for finding an answer to our question.

The problem of the identification of the solar minimum/maximum moments in the Maunder Minimum period was solved by different methods (Usoskin *et al.*, 2001). For each of these methods the moments were obtained. Unfortunately, the moments obtained by different methods are in low agreement one with another. We will not discuss here the specific problems of each of the methods because it is outside the framework of this paper. We note only that our choice of the $^{10}$Be from Greenland ice as a source of solar activity data was made because of the direct connection between $^{10}$Be and the cosmic ray flux. The relatively close proximity of Greenland to England was an additional reason for this choice. The source of the data on $^{10}$Be is an article of Prof. Beer *et al.* (1998), on which the original graph of $^{10}$Be variation from year to year was published. The results of applying the spectral pass filters, 7-24 years and 9-14 years, to the original $^{10}$Be data were also used. To identify minimums and maximums in the sunspot activity, Prof. Beer used the graph obtained after applying the intermediate filter (7-24 years). For our purpose, we used the results of Prof. Beer on cycle variability, obtained after narrower spectral pass filter of 9-14 years. We chose this filter because this interval is closer to our observed interval of variability periods and the data is minimally distorted by variations of another nature with another spectrum. In the Appendix, the revised identification of the maximum/minimum moments is presented.

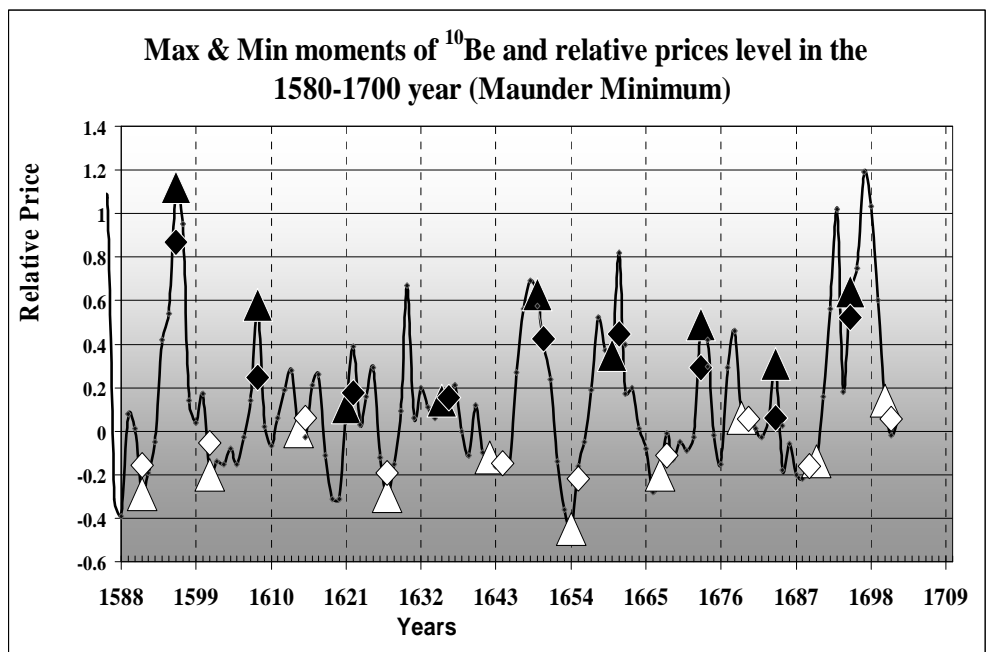

**Figure 9.** Consistent differences in prices at moments of maximum and minimum states of solar activity (1600-1700). White and black rectangles are prices averaged for 3-years intervals centered on moments of maximum and minimum of solar activity, white and black triangles are prices in the moment of the maximum and minimum.

In Fig. 9 the price variations for the century of the Maunder Minimum, 1600-1700, are presented. On the graph, white rectangles represent average prices at 3-year intervals around the moments of the minimal $^{10}$Be contents (maximum solar activity), taken from the tables of Beer *et al.* (1998), based on data filtered 7-24 years, and Usoskin *et al.* (2001). Uncertainty concerning the time lag between $^{10}$Be and solar activity (+/- 1 year) forced us to use these three-year averages for higher reliability of the choice. The intervals of the minimal solar activity were defined as three-year intervals around moments which were exactly between two adjacent moments of maximum solar activity. The corresponding average prices of these three-year intervals of minimal activity are marked by black rectangles on the graph.

The prices of the selected moments of the maximal $^{10}$Be are marked by white triangles, and the moments of minimal $^{10}$Be are marked by black triangles. These time moments were obtained from the digitized graph of Beer *et.al.*, Fig. 2b (1998), based on data filtered 9-14 years (Appendix, Fig.A.2.).

As it can be seen from Fig. 9, all wheat prices in the phases of minimum solar activity (maximum $^{10}$Be and cosmic rays input) are consistently higher than the corresponding prices in the phases of maximum solar activity (minimal $^{10}$Be and cosmic ray intensity). This conclusion is true both for three-year averages from the Beer data filtered 7-24 years (rectangles) and for maximum/minimum moments from our analysis of Beer data filtered 9-14 years (triangles). The probability of a random occurrence of this consistent difference may be estimated through the criterion of the sign correlation as **W = (1/2)$^9$ < 0.2%.** In this case, any explanation of closeness between two different cyclical processes (wheat price variations and solar spot variations) by random coincidence of the periods is unacceptable, since the period of the solar cycle varies strongly.

Additionally, it should be noted that amplitudes of the price differences between those in minimum and in corresponding maximum solar activity states (triangles) average about 70%, which is twice that of the differences for averaged 3-year intervals (about 30%).

## 6. Discussion and Conclusion

The results of our study show:

a) The coincidence between the statistical properties of the distributions of intervals between wheat price bursts in medieval England (1259-1702) and intervals between minimums of solar cycles (1700-2000);

b) The existence of 100% sign correlation between high wheat prices and states of minimal solar activity established on the basis of $^{10}$Be data for Greenland ice measurements for the period 1600-1700.

These results imply a causal connection between solar activity and wheat prices in medieval England. This conclusion is consistent with our conceptual model of the causal chain, consisting of "solar activity – cosmic ray intensity - terrestrial weather – wheat production – wheat price" that presented in this work. The sign of the causal connection (maximum prices for minimum sunspot activity and maximum cosmic ray flux) is the same sign that can be expected for agriculture/weather conditions in medieval England. We reiterate that England, at that time, was a region of "high risk" agriculture resulting from the "cold" side of the weather axis, when "…the cultivation of wheat was not carried on successfully beyond the north bank of the Humber" (Rogers, 1887, V.1, p. 29).

Regarding possible manifestations of solar activity in the dynamics of prices of agricultural production in modern times, we believe that the causality shown in the above causal chain "solar activity - … - wheat price" should remain valid. More generally, one can think of cereal production and prices in this chain. Clearly, however, the globalization of international food markets by transcontinental shipping leads to diminished local effects on variations of these prices. Another stabilizing factor results from successful development and implementation of genetic, selective and agrochemical technologies, which can widen feasible conditions for cereals production (zones "**b**", "**c**", "**d**" in Fig. 3). However, the last

1650 years of global warming could lead to shifts in climatic conditions for different regions, especially near desert areas, to "boundary" conditions, making them more sensitive to hot summers and droughts, with catastrophic drops in agricultural production. As the result of the climatic changes and other negative factors of socio-economic nature, the danger of famine in developing countries placed in the high-risk agriculture zones is still very real. In these countries, the influence of variations in solar activity could be essential for agricultural production.

On the other hand, large-scale international activities are devoted to help these countries. The aid of the international community comes in the form of support in agricultural research and reforms in these countries, and in the form of direct food supplies in extreme periods. Discovery of the manifestation of the solar activity in cereal production and prices could supply important information for planning the expected need for direct food supplies, and for the agro-economic evaluation of different crops in zones of high-risk agriculture.

## Acknowledgements

We thank Prof. L.Dorman, Prof. G.Beer and Prof. L. Alperovich for helpful discussions.

**APPENDIX. Identification of maximum/minimum from Beer *et al*. (1998) data.**

As was shown by Beer *et al*. (1998) $^{10}$Be variations in Greenland ice involve different components, from short-term (one-year) bursts and slow systematic trends up to periodical variances over a time of about 10+/-3 years (see Fig. A1.a).

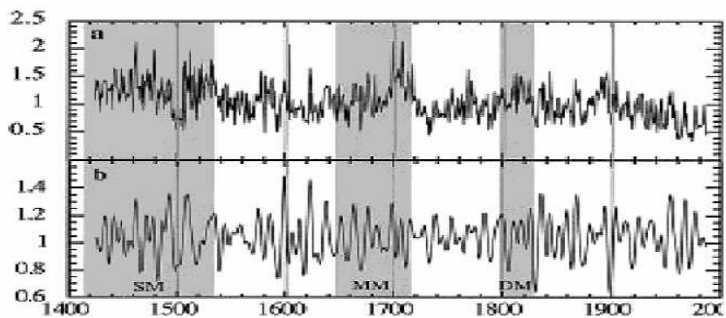

Figure A1: a) Original data of Beer et al., 1998/; b) the same after frequency band filter 7-28 days

| Table A1. Minimums and maximums of $^{10}$Be, 1580-1720 years | |
|---|---|
| **Minimums of $^{10}$Be from 1580-1720 filtered 9-14 years** | |
| Year | Content $^{10}$Be |
| 1601.8 | 0.84 |
| 1614.5 | 0.87 |
| 1627.3 | 0.93 |
| 1641.8 | 0.94 |
| 1654.5 | 0.88 |
| 1666.8 | 0.87 |
| 1679.5 | 0.87 |
| 1690 | 0.92 |
| 1700 | 0.99 |
| 1709.1 | 0.93 |
| 1719.5 | 0.90 |
| **Maximums of $^{10}$Be from 1580-1720 filtered 9-14 years** | |
| 1595.9 | 1.27 |
| 1597.7 | 2.44 |
| 1608.6 | 1.24 |
| 1620.9 | 1.20 |
| 1621.4 | 2.44 |
| 1634.5 | 1.13 |
| 1648.6 | 1.21 |
| 1660.4 | 1.23 |
| 1673.1 | 1.22 |
| 1684.5 | 1.19 |
| 1695 | |

To clean the search cyclical process from other components of $^{10}$Be variations, Beer *et al. (1998)* applied two spectral pass filters: 7-24 years and 9-14 years. For the next identification, these authors used min/max point of the filtered data with a wide filter - 7-24 year (high curve in Fig. A2.a) - which results in not reliable enough peaks and disturbances. Their resulting minimum/maximum moments were shown in Table 1 of Beer *et al*. (1998).

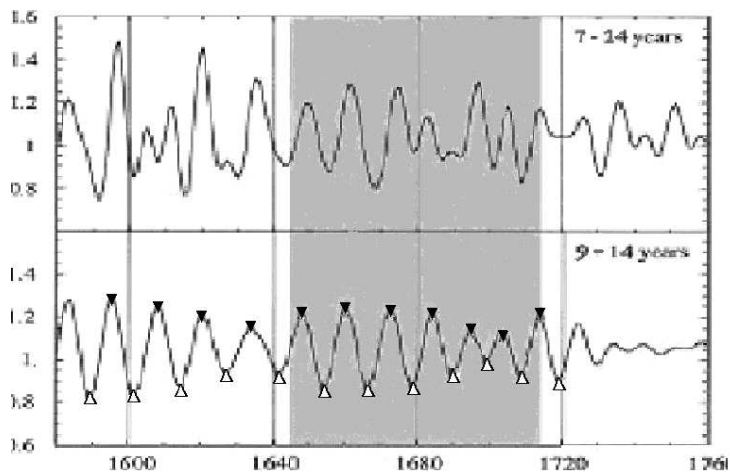

Figure A2. Filtered data of Beer et al., 1998 after applying 7-24 years and 9-14 years filters with marked point (triangles) of the minimum and maximum identification in the low graph for Table A1.

Filtered data after applying filters 7-24 years (a) and 9-14 years (b) for the time interval of the Maunder Minimum 1580-1720 years are shown in Fig. A2. The last curve (b), after digitizing, is used for the identification of the minimum and maximum states of $^{10}$Be content, corresponding to maximum and minimal states of sunspot cycle activity. White triangles mark minimum states of $^{10}$Be and black triangles mark maximum states. The resulting moments of maximums and minimums of $^{10}$Be are presented in Table A1.